
\input phyzzx

\REF\BST{E. Bergshoeff, E. Sezgin and P.K. Townsend, Phys. Lett. {\bf
189B}  (1987) 75;
Ann. Phys. (N.Y.) {\bf 185} (1988) 330.}
\REF\DHIS{M.J. Duff, P.S. Howe, T. Inami and K.S. Stelle, Phys. Lett.
{\bf 191B} (1987) 70.}
\REF\SS{J.H. Schwarz and A. Sen, Nucl. Phys. {\bf B411} (1994) 35;
Phys. Lett. {\bf 312B} (1993) 105.}
\REF\Sen{A. Sen, Nucl. Phys. {\bf B404} (1993) 109; Phys. Lett. {\bf
303B} (1993).}
\REF\HT{C.M. Hull and P.K. Townsend, {\it Unity of superstring
dualities}, Nucl. Phys. {\bf B} {\sl in press}.}
\REF\BP{I. Bars, C.N. Pope and E. Sezgin, Phys. Lett. {\bf 198B} (1987) 455.}
\REF\ST{K.S. Stelle and P.K. Townsend, {\it Are 2-branes better than
1?} in Vol. 2 of {\sl Field theory in two dimensions, and related
topics}, eds. G. Kunstatter, H.C. Lee, F.C. Khanna and H. Umezawa
(World Scientific 1988);
P.K. Townsend, {\it Three lectures on supermembranes} in {\sl
Superstrings '88}, eds. M. Green, M. Grisaru, R. Iengo, E. Sezgin and
A. Strominger (World Scientific 1989).}
\REF\dWNL{B. de Wit, M. L\"uscher and H. Nicolai, Nucl. Phys. {\bf
B320} (1989) 135.}
\REF\DS{M.J. Duff and K.S. Stelle, Phys. Lett. {\bf 253B} (1991) 113.}
\REF\DGT{M.J. Duff, G.W. Gibbons and P.K. Townsend, Phys. Lett. {\bf
332 B} (1994) 321.}
\REF\HS{G.T. Horowitz and A. Strominger, Nucl. Phys. {\bf B360} (1991) 197.}
\REF\NT{R. Nepomechie, Phys. Rev. {\bf D31} (1985) 1921; C.
Teitelboim, Phys. Lett. {\bf B167} (1986) 69.}
\REF\DR{M.J. Duff and J. Rahmfeld, {\it Massive string states as
extreme black holes}, preprint CTP-TAMU-25/94.}
\REF\Susskind{L. Susskind, {\it Some speculations about black hole
entropy in string theory}, preprint hep-th/9309145.}
\REF\GHT{G.W. Gibbons, G.T. Horowitz and P.K. Townsend, {\it
Higher-dimensional resolution of dilatonic black hole singularities},
Class. and Quantum Grav. {\sl in press}.}
\REF\KKM {R. Sorkin, Phys. Rev. Lett. {\bf 51} (1983) 87; D. Gross and
M. Perry, Nucl. Phys. {\bf B226} (1983) 29.}
\REF\DGHR{A. Dabholkar, G.W. Gibbons, J.A. Harvey and F. Ruiz-Ruiz,
Nucl. Phys. {\bf B340} (1990) 33.}
\REF\G{R. G\"uven, Phys. Lett. {\bf 276B} (1992) 49.}
\REF\GT{G.W. Gibbons and P.K. Townsend, Phys. Rev. Lett. {\bf 71} (1993) 3754.}

\font\mybb=msbm10 at 12pt
\def\bb#1{\hbox{\mybb#1}}
\def\R {\bb{R}}

\def\square{\kern1pt\vbox{\hrule height 0.5pt\hbox
{\vrule width 0.5pt\hskip 2pt
\vbox{\vskip 4pt}\hskip 2pt\vrule width 0.5pt}\hrule height
0.5pt}\kern1pt}


\Pubnum{ \vbox{ \hbox{R/95/2} \hbox{9501068}} }
\pubtype{}
\date{January, 1995}

\titlepage

\title{The eleven-dimensional supermembrane revisited}

\author{P.K. Townsend}
\address{DAMTP, Univ. of Cambridge, Silver St., Cambridge, U.K. }

\abstract {It is argued that the type IIA 10-dimensional superstring
theory is actually a compactified 11-dimensional supermembrane theory
in which the fundamental supermembrane is identified with the the
solitonic membrane of 11-dimensional supergravity. The charged extreme
black holes of the 10-dimensional type IIA string theory are
interpreted as the Kaluza-Klein modes of 11-dimensional supergravity
and the dual sixbranes as the analogue of Kaluza-Klein monopoles. All
other p-brane solutions of the type IIA superstring theory are derived
from the 11-dimensional membrane and its magnetic dual fivebrane soliton.}

\endpage

\pagenumber=2



The effective field theory of the ten-dimensional type IIA superstring is N=2A
supergravity. It has long been appreciated that this field theory is
also the effective massless theory for eleven-dimensional supergravity
compactified on $S^1$; the ten-dimensional dilaton thereby acquires a
natural Kaluza-Klein (KK) interpretation. This leads one to wonder
whether the type IIA string theory has an eleven-dimensional
interpretation. An obvious candidate is the 11-dimensional
supermembrane [\BST] since the double dimensional reduction of its
worldvolume action yields the Green-Schwarz (GS) action of the type
IIA superstring [\DHIS]. Despite this, the 11-dimensional
interpretation of the {\it quantum} type IIA superstring is obscure
because the dilaton vertex operator is radically different from the
graviton vertex operator. In the GS action the dilaton comes from the
R-R sector while the graviton comes from the NS-NS sector; there is
therefore no obvious KK interpretation of the dilaton in string theory
(in the bosonic string the dilaton is usually taken to couple to the
worldsheet curvature but this makes the dilaton vertex operator even
more dissimilar to the graviton vertex operator). It is possible,
however, that this special status of the dilaton is an artefact of
perturbation theory. It has recently been realized that some features
of the effective field theories of compactified superstring theories,
such as invariance under a generalized electromagnetic duality, may
also be features of the full non-perturbative string theory even
though this is not apparent in perturbation theory [\SS,\Sen,\HT]. In
this letter I similarly argue that the type IIA 10-dimensional
superstring theory actually {\it is} a compactified 11-dimensional
supermembrane theory.

Before further analysis of this conjecture, some discussion of the
status of the 11-dimensional supermembrane is warranted. There is good
reason to suppose that the supermembrane spectrum contains massless
particles  which can be identified as the graviton and other quanta of
11-dimensional supergravity [\BP]. The principal objection to this
conclusion  is that there are also reasons [\ST,\dWNL] to believe the
spectrum  to be continuous, which would preclude a particle
interpretation.  The physical reason for this is that there is no
energy cost  to a deformation of the membrane leading to `spikes' of
arbitrary length but zero area, like a fakir's bed of nails (for the
bosonic  membrane there is an energy cost at the quantum level due to
the  Casimir effect, but this Casimir energy cancels for the
supermembrane).  The possibility of spikes of zero
area is of course due to the supposition that the membrane has a core
of zero  width. A calculation [\dWNL] in the context of a
first-quantized,  regularized, zero-width supermembrane showed that
the  spectrum is indeed continuous, from zero, and this was widely
interpreted  as putting an end to the idea of a `fundamental' supermembrane.

However, evidence was presented in [\HT] that the fundamental
supermembrane  should be identified with the solitonic membrane [\DS]
of  11-dimensional supergravity. An additional reason for this
identification  is that $\kappa$-symmetry of the worldvolume action
for a  supermembrane requires the background fields to satisfy the
{\it  source-free} field equations of 11-dimensional supergravity
[\BST].  This is paradoxical if the supermembrane is regarded as the
source  of the background fields, but the paradox would be resolved if
the  fundamental supermembrane were to be identified with a membrane
solution  of the source-free field equations, and the one of [\DS] is
the only candidate. As originally presented this was seen as the
exterior  solution to a singular surface, which was interpreted as a
membrane  source, but the singularity can be interpreted equally well
as a  mere coordinate singularity at an event horizon, through which
the  source-free exterior solution can be analytically continued
[\DGT]. If  one accepts the identification of the fundamental and
solitonic  supermembranes in the fully non-perturbative quantum
theory, then  it follows that the supermembrane acquires a core of
finite size  due to its gravitational field in the same way that a
`point'  particle actually has a size of the order of its
Schwarzschild radius  once gravitational effects are included. In this
case a  `spike' of a given length has a minimum area and therefore a
minimum  energy cost. Under these circumstances one would not expect a
continuous spectrum. A possible objection to this argument is that it
could  also be applied to string theory where, however, it is not
needed  because the spectrum is already discrete in perturbation
theory. This  may simply be a reflection of the fact that perturbation
theory  makes sense for strings because of the renormalizability of
two-dimensional sigma-models whereas it does not make sense for
membranes  because of the non-renormalizability of three-dimensional
sigma  models. In any case, I shall assume in the following that the
fully non-perturbative supermembrane spectrum is discrete for reasons
along  the above lines. It is perhaps worth mentioning here that a
similar  argument would be needed to make sense of a 10-dimensional
`fundamental' fivebrane, so that evidence in favour of
string-fivebrane  duality can be construed as evidence that
non-perturbative  effects cause the spectrum of a `fundamental'
fivebrane to  be discrete, and if this is case for fivebranes then why
not for  p-branes in general?

The determination of the spectrum of the 11-dimensional supermembrane,
given  that it is discrete, is impossible in practice, as it is for
superstrings when account is taken of interactions and all
non-perturbative  effects. However, certain features of the spectrum
can be  reliably ascertained. Among these is the massless spectrum,
for which  the effective field theory is just 11-dimensional
supergravity.  This theory reduces to 10-dimensional N=2A supergravity
upon  compactification on $S^1$, but the spectrum in 10-dimensions
will then  also include the charged massive KK states. These states
must also  be present in the spectrum of the type IIA superstring if
the  latter is to be interpreted as a compactified supermembrane, as
conjectured here. These states do not appear in perturbation theory
but there  are extreme black hole solutions of 10-dimensional N=2A
supergravity that are charged with respect to the KK $U(1)$ gauge
field [\HS].  Because these solutions preserve half of the
supersymmetry there  are good reasons (see e.g. [\HT] and references
therein)  to believe that their semi-classical quantization of will be
exact.  I suggest that these states be identified as KK states. I
shall now  address  possible objections to this identification.

First, the mass of a KK state is an integer multiple of a basic unit
(determined by the $S^1$ radius) whereas the mass of an extreme black
hole is  apparently arbitrary. However, there are also 6-brane
solutions of  N=2A supergravity [\HS] that are the magnetic duals of
the  extreme black holes. It will be shown below that these 6-branes
are  completely non-singular when interpreted as solutions of the
compactified  11-dimensional supergravity. It follows, if the
11-dimensional  interpretation is taken seriously, that the 6-brane
solitons  must be included as solutions of the ten-dimensional theory
and  then, by the generalization of the Dirac quantization condition
to  p-branes and their duals [\NT], we conclude that in the quantum
theory the  electric charge of the extreme black holes is quantized.
Since  their mass is proportional to the modulus of their charge, with
a  universal constant of proportionality, their mass is also
quantized. The  {\it unit} of mass remains arbitrary, as was the $S^1$ radius.

Second, it may be objected that whereas the type IIA theory has only
one set  of charged states coupling to the $U(1)$ gauge field, the
compactified supermembrane theory has {\it two}: the extreme black
hole  solutions of the effective 10-dimensional field theory after
compactification on $S^1$ {\it and} the KK modes. The two sets of
states have  identical quantum numbers since the allowed charges must
be the  same in both cases. It has recently been argued in the context
of  compactifications of the heterotic [\DR] and the type II [\HT]
superstrings that KK states should be {\it identified} with
electrically  charged extreme black holes (see also [\Susskind]). The
reasons  advanced for this identification do not obviously apply in
the  present context but once the principle is granted that this
identification is possible it seems reasonable to invoke it more
generally.  Thus, I conjecture that the resolution of this second
objection  is that the KK and extreme black hole states of the $S^1$
compactified 11-dimensional supergravity are not independent in the
context of  the underlying supermembrane theory. This conjecture is
similar to  those made recently for the heterotic and type II
superstrings  but there is a crucial difference; in the string theory
case the  KK states also appear in the perturbative string spectrum since they
result from compactification {\it from} the critical dimension,
whereas the KK  states discused here do not appear in the perturbative
string  spectrum because they result from compactification {\it to}
the  critical dimension.

Little more can be said about the spectrum of {\it particle} states in
ten  dimensions since only those solutions of the effective field
theory that  do not break all supersymmetries can yield reliable
information  about the exact spectrum upon semi-classical
quantization, and  the only such particle-like solutions are the
extreme electric black holes. However, there are also p-brane solitons
of N=2A  supergravity which preserve half the supersymmetry and are
therefore  expected to be exact solutions of type IIA string theory.
These  should also have an 11-dimensional interpretation. The 6-brane
soliton  has already been mentioned; we now turn to its 11-dimensional
interpretation. Consider the 11-metric
$$
ds^2_{11} = -dt^2 + dy\cdot dy + V({\bf x}) d{\bf x}\cdot d{\bf x} +
V^{-1}({\bf x}) \big(dx^{11} -{\bf A}({\bf x})\cdot d{\bf x}\big)^2\ ,
\eqn\one
$$
where $dy\cdot dy$ is the Euclidean metric on $\R^6$ (an infinite
planar  6-brane) and $d{\bf x}\cdot d{\bf x}$ is the Euclidean metric
on  $\R^3$ (the uncompactified transverse space). This metric solves
the  11-dimensional vacuum Einstein equations, and hence the field
equations  of 11-dimensional supergravity when all other fields are
set to  zero, if $\nabla\times {\bf A} =\nabla V$, which implies that
$\nabla^2 V=0$. One solution is
$$
V= 1+ {\mu\over\rho}
\eqn\two
$$
where $\rho=\sqrt{{\bf x}\cdot {\bf x}}$ and $\mu$ is a constant. The
two-form  $F=dA$ is then given by
$$
F=\mu\varepsilon_2\ ,
\eqn\three
$$
where $\varepsilon_2$ is the volume form on the unit 2-sphere. The
singularity  at $\rho=0$ is merely a coordinate singularity if
$x^{11}$ is  identified modulo $4\pi\mu$. Thus \one\ is a {\it
non-singular}  solution of compactified 11-dimensional supergravity
representing a magnetic KK 6-brane. It is an exact
analogue in 11 dimensions of the KK monopole in 5 dimensions [\KKM].
Considered as a solution of the effective field theory of
ten-dimensional  string theory, the 10-metric, in `string conformal gauge', is
$$
ds^2_{10} = \Big( 1+{\mu\over\rho}\Big)^{-{1\over2}}\Bigg[ -dt^2 +
dy\cdot dy  + \Big(1+{\mu\over\rho}\Big)d{\bf x}\cdot d{\bf x}\Bigg]
\eqn\four
$$
while the 10-dimensional dilaton field $\phi$ is given by
$$
e^{-2\phi} = \Big(1+{\mu\over\rho}\Big)^{3\over2}\ .
\eqn\five
$$
In terms of the new radial coordinate $r=\rho +\mu$, we have
$$
\eqalign{
ds^2_{10} &= \Big(1-{\mu\over r}\Big)^{1\over2} \Big[ -dt^2 + dy\cdot
dy\Big]  + \Big(1-{\mu\over r}\Big)^{-{1\over2}} dr^2 + r^2
\Big(1-{\mu\over  r}\Big)^{3\over2} d\Omega_2^2 \cr
e^{-2\phi} &= \Big(1-{\mu\over r}\Big)^{-{3\over2}}\ ,}
\eqn\six
$$
where $d\Omega_2^2$ is the metric on the unit 2-sphere. This
is just the 6-brane solution [\HS] of 10-dimensional N=2A supergravity.

The remaining p-brane soliton solutions of N=2A supergravity are the
string  [\DGHR], membrane, fourbrane and fivebrane [\HS]. The string
and  fourbrane solitons have previously been shown [\DHIS,\DGT] to be
double-dimensional reductions of, respectively, the 11-dimensional
membrane  and the 11-dimensional fivebrane [\G]. The 10-dimensional
membrane  and fivebrane differ from their 11-dimensional counterparts
simply  by the boundary conditions imposed on the solution of the
Poisson  equation with a point source that always arises in the
context of the  extreme p-brane solitons, and the ten-dimensional
soliton can  be viewed as a periodic array of 11-dimensional solitons.
Thus,  all p-brane solitons of 10-dimensional N=2A supergravity have
an  11-dimensional origin. Moreover, since the 11-dimensional
fivebrane has a  completely non-singular analytic extension through
its  horizon [\GHT], the 10-dimensional magnetic 4,5 and 6-brane
solitons are  all {\it completely non-singular} when interpreted as
solutions  of compactified 11-dimensional supergravity. This can be
taken as  further evidence in favour of an 11-dimensional origin of
these  solutions of the apparently 10-dimensional type IIA superstring
theory.  It is perhaps worth remarking that, not surprisingly, there
is no  similar interpretation of the p-brane solitons of type IIB
superstring  theory.

It may be objected here that while all of the p-brane {\it solitons}
of the  type IIA superstring may be solutions of an $S^1$-compactified
supermembrane theory, the two theories differ in that one has an
additional  fundamental string while the other has an additional
fundamental  membrane. But this difference disappears once one
identifies the  fundamental string or membrane with the solitonic
ones; both  theories then have exactly the same spectrum of extended
objects.  In fact, it becomes a matter of convention whether one calls
the  theory a string theory, a membrane theory, or a p-brane theory
for any  of the other values of p for which there is a soliton
solution; all  are equal partners in a p-brane democracy.

In 11-dimensions the fivebrane soliton is the magnetic dual of the
11-dimensional membrane soliton, which has been identified with a
fundamental  supermembrane. As suggested in [\HT], one can envisage a
dual  11-dimensional fivebrane theory in which the soliton fivebrane
is  identified with a fundamental 11-dimensional fivebrane. Since the
physical  (i.e. gauge-fixed) worldvolume action of the latter must be
[\GT] a  {\it chiral} six-dimensional supersymmetric field theory
based on the  self-dual antisymmetric tensor supermultiplet, it seems
possible  not only that there is a consistent quantum theory of
11-dimensional  supergravity based on the supermembrane, but also that
its  dual formulation might lead to a solution to the chirality
problem that  bedevils any attempt to obtain a realistic model of
particle  physics starting from 11-dimensions.

\vskip 1cm
\centerline{\bf Acknowledgments}
Discussions with M.J. Duff, C.M. Hull and K.S. Stelle are gratefully
acknowledged

\refout

\end